\documentclass[journal=aelccpcmatex,manuscript=article]{achemso}
\usepackage[version=4]{mhchem}
\usepackage{bm}
\usepackage{amsmath}
\usepackage{amsfonts}
\usepackage{amssymb}
\usepackage{graphicx}
\usepackage{setspace}
\usepackage{lineno}
\usepackage{color}
\usepackage{comment}

\author{Yonghyuk Lee}
\affiliation{Department of Chemistry and Biochemistry, University of California Los Angeles, Los Angeles, California, 90095, United States}
\author{Taehun Lee}
\email{taehun.lee@jbnu.ac.kr}
\affiliation{Division of Advanced Materials Engineering, Jeonbuk National University, Jeonju 54896, Republic of Korea}
\alsoaffiliation{Hydrogen and Fuel Cell Research Center, Jeonbuk National University, Jeonbuk, 54896, Republic of Korea}

\title{Machine-Learning-Accelerated Surface Exploration of Reconstructed \ce{BiVO4}(010) and Characterization of Their Aqueous Interfaces}

\begin{document}


\begin{abstract}
Understanding the semiconductor-electrolyte interface in photoelectrochemical (PEC) systems is crucial for optimizing stability and reactivity. Despite the challenges in establishing reliable surface structure models during PEC cycles, this study explores the complex surface reconstructions of \ce{BiVO4}(010) by employing a computational workflow integrated with a state-of-the-art active learning protocol for a machine-learning interatomic potential and global optimization techniques. Within this workflow, we identified 494 unique reconstructed surface structures that surpass conventional chemical intuition-driven, bulk-truncated models. After constructing the surface Pourbaix diagram under Bi- and V-rich electrolyte conditions using density functional theory and hybrid functional calculations, we proposed structural models for the experimentally observed Bi-rich \ce{BiVO4} surfaces. By performing hybrid functional molecular dynamics simulations with explicit treatment of water molecules on selected reconstructed \ce{BiVO4}(010) surfaces, we observed spontaneous water dissociation, marking the first theoretical report of this phenomenon. Our findings demonstrate significant water dissociation on reconstructed Bi-rich surfaces, highlighting the critical role of bare and under-coordinated Bi sites (only observable in reconstructed surfaces) in driving hydration processes. Our work establishes a foundation for understanding the role of complex, reconstructed Bi surfaces in surface hydration and reactivity. Additionally, our theoretical framework for exploring surface structures and predicting reactivity in multicomponent oxides offers a precise approach to describing complex surface and interface processes in PEC systems.
\end{abstract}

\maketitle

\section{INTRODUCTION}

Understanding semiconductor-electrolyte interfaces (SEI) in photoelectrochemical (PEC) systems is inherently challenging, particularly during operational conditions~\cite{Monllor2020}.
The difficulty in studying these interfaces has historically limited our understanding, often relying on \textit{ex-situ} data or theoretical models.
However, recent advancements in \textit{operando} characterization techniques now allow real-time tracking of SEI, providing direct insights into interfacial processes~\cite{Barawi2024}.
These techniques have unveiled complex surface reconstructions in multicomponent metal oxides such as \ce{BiVO4} during the PEC illumination cycle~\cite{Venugopal2021}.

\ce{BiVO4} interfaces, in particular, have been shown to undergo preferential vanadium (V) dissolution under open-circuit conditions, while both bismuth (Bi) and vanadium dissolve under anodic conditions, altering the Bi/V ratio at the surface over time~\cite{Lee2021,Venugopal2021}.
Despite these findings, the precise structures and local atomic geometries of the reconstructed Bi-rich (or Bi/V mixed) surface layers during the PEC cycle remain unclear.
Understanding these structures is crucial for comprehending surface energetics and dynamics at aqueous interfaces for optimizing PEC efficiency.

The complex reconstruction and stability of multicomponent surfaces have been extensively studied using \textit{ab initio} calculations, particularly density functional theory (DFT).
DFT complements experimental data by providing detailed atomic structures and insights into their thermodynamic stability.
However, predicting surface structures based solely on experimental data or chemical intuition is limited, especially in identifying global minima on the potential energy surface (PES).
This approach often depends heavily on initial structural guesses, which may not fully capture the complexity of surface processes~\cite{Reuter2016,Lee2024}.
To address these limitations, global optimization algorithms, such as simulated annealing, basin hopping, and genetic algorithms~\cite{Kresse2003,Vilhelmsen2014,Panosetti2015,Wexler2019,Zhou2019,Zhang2022}, have been coupled with DFT to explore the PES and identify global minima.
While successful, these approaches require extensive computational resources to identify stable surface structures across various chemical compositions and stoichiometries.

Machine-learning interatomic potentials (MLIPs) offer a promising solution by reducing computational burdens.
By training MLIPs on a limited set of high-cost DFT calculations, a reliable surrogate PES can be constructed, retaining DFT’s predictive accuracy~\cite{Deringer2019,Willatt2019,Zuo2020,Deringer2021,Musil2021,Kocer2022}.
This enables global optimization techniques to be conducted on low-cost MLIPs, significantly extending simulation timescales and system sizes.
Moreover, emerging active-learning strategies enhance the accuracy of the surrogate PES by integrating the generation of new training data with global optimization processes, ensuring it closely aligns with the underlying DFT predictions in relevant domains while maximizing data efficiency~\cite{Timmermann2021,Lee2023}.

Despite these advancements, theoretical exploration of surface structures in multicomponent oxides under aqueous electrochemical (EC) conditions remains incomplete and is rarely investigated.
This gap is largely due to the increased complexity of the electrochemical surface processes in comparison to gas-solid systems in the vacuum condition.
Comparing surface stability under EC conditions requires additional thermodynamic considerations for pH and electrode potential, alongside the incorporation of experimental formation energies for water and ionic species in electrolytes.
Recent studies suggest that accurately constructing Pourbaix diagrams for semiconducting systems often demands methods beyond standard DFT, such as meta-GGA or high-cost hybrid functionals~\cite{Zeng2015,Huang2019}.

In this study, we reveal the complex surface reconstruction behavior of \ce{BiVO4}(010) and the aqueous interfacial properties of these reconstructed surfaces.
We employ a multi-stage approach, as outlined in Figure~\ref{fig1}, encompassing (i) \textit{surface structure exploration}, (ii) \textit{EC stability evaluations}, and (iii) \textit{predictions of aqueous interfacial properties}, with each stage refining the systems of interest.

(i) \textit{Surface structure exploration stage}: We first developed a cost-efficient MLIP surrogate model based on \textit{ab initio} training data for \ce{BiVO4}(010), which was consistently refined using a state-of-the-art, data-efficient MLIP training protocol within an active-learning workflow.
This protocol combines the exploration of surface structures by navigating compositional and configurational spaces with a global optimization technique, i.e., simulated annealing (SA), with on-the-fly training of MLIP.
Through the SA cycle, which allows for lateral mass transfer of surface species during heating and identifies (meta)stable configurations during cooling, the complex surface reconstructions on \ce{BiVO4}(010) were captured, resulting in a total of 494 unique structures.
(ii) \textit{EC stability evaluations}: In the next stage, EC stability was evaluated to understand the experimentally observed surfaces (yet not reported for exact structures) using a two-step DFT screening process with standard DFT using the Generalized Gradient Approximation and cost-efficient dielectric-dependent hybrid functional calculations.
We then refined our list to 6 structures, stable under Bi- and V-rich electrolyte and PEC reactive conditions, whose geometries extend beyond the simple chemical intuition-driven, bulk-truncated \ce{BiVO4}(010).
(iii) \textit{Predictions of aqueous interfacial properties}: We also performed \textit{ab initio} molecular dynamics simulations for the aqueous interface of reconstructed \ce{BiVO4}(010) surfaces using dielectric-dependent hybrid functionals to analyze their interfacial properties.
To our knowledge, this is the first theoretical study to report spontaneous water dissociation via direct and indirect proton transfer on \ce{BiVO4}(010).
The reconstructed \ce{BiVO4}(010) surfaces, both Bi- and V-rich, exhibited significant water dissociation compared to the stoichiometric surface.
The Bi-rich surfaces, in particular, showed a greater degree of water dissociation, attributed to the presence of under-coordinated Bi atoms.
Altogether, our findings suggest that the complex, reconstructed Bi surfaces can enhance surface hydration, potentially explaining the PEC-active structures observed in recent experiments.

We anticipate that our predicted surface structures will be validated by future experimental studies.
Our approach, which integrates MLIP-MD with cost-effective hybrid functional calculations, provides a robust framework for exploring complex surface processes and deepening our understanding of (multicomponent) semiconductor-electrolyte interfaces, ultimately advancing the development of more efficient PEC systems.

\section{METHOD}
\subsection{DFT Calculations.}

The training data for developing the Gaussian Approximation Potential (GAP) MLIP model for \ce{BiVO4}(010) was generated using DFT calculations, employing optimized norm-conserving Vanderbilt pseudopotentials (ONCVPSP~\cite{Hamann2013}) within the Quantum ESPRESSO package~\cite{Giannozzi2009}.
We used the Generalized Gradient Approximation (GGA) with the Perdew-Burke-Ernzerhof (PBE) functional~\cite{Perdew1996}, applying a kinetic energy cutoff of 90~Ry for the wave function expansion and a charge density cutoff of 360~Ry.
Brillouin-zone integrations were conducted using a k-point grid with reciprocal distances of 0.05~{\AA}$^{-1}$, corresponding to a ($6 \times 6 \times 4$) grid for primitive bulk \ce{BiVO4}.

For simplicity, we utilized the tetragonal phase of \ce{BiVO4} (space group $I4_{1}/a$) rather than the room-temperature monoclinic phase (space group $C2/c$), as both phases exhibit similar band structures and electron polaronic states~\cite{Seo2018}.
The tetragonal (001) surface is symmetrically similar to the (010) plane of the monoclinic phase, so we refer to the investigated (001) facet of the tetragonal phase as the (010) surface. We adopted the lattice parameters ($a=5.172$ and $c=11.771$~{\AA}) of bulk \ce{BiVO4} from Ref.~\citenum{Seo2018} for benchmarking and comparison with related studies~\cite{Wang2020,Lee2021,Hilbrands2023}. 
The \ce{BiVO4}(010) slab was constructed as a symmetric slab model with periodic boundary conditions and a minimum vacuum separation of 20~{\AA}.
The model comprised at least ten trilayers of \ce{BiVO4} within the $(1 \times 1)$ surface unit cell.

Constrained geometry optimizations of the initial or during active learning \ce{BiVO4}(010) slab models were performed using the Broyden-Fletcher-Goldfarb-Shanno (BFGS) minimization algorithm~\cite{Broyden1970,Goldfarb1970,Shanno1970}, with convergence criteria set to total energy below $1.4\times10^{-2}$~meV and force components below 0.3~meV/{\AA}.
To efficiently capture the bonding characteristics of the slab models during active learning, we intentionally terminated the optimizations within 20 ionic steps.

After exploring the unique surfaces using GAP, we re-optimized the obtained \ce{BiVO4} surfaces using DFT until the energy and forces met the convergence criteria. 
To achieve an accurate description of the electrochemical stability of the reconstructed \ce{BiVO4} surfaces, we further performed geometric optimization of selected slab models using dielectric-dependent PBE0 hybrid functional calculations.
In this stage, we extended our $p(1\times1)$ structures to the $p(2\times2)$ supercell, allowing for potential reconstructions due to the higher degree of freedom.
The Brillouin zone was sampled at the $\Gamma$-point for supercell calculations.

The PBE0 calculations, with the fraction of exact exchange ($\alpha$) set to $\alpha = 1/\epsilon_\infty = 0.22$, following a previous computational setup~\cite{Wiktor2017,Ambrosio2019}, were accelerated using the Truncated Coulomb and Long Range Correction (PBE0-TC-LRC) method~\cite{Guidon2009} along with the auxiliary density matrix method~\cite{Guidon2010}, as implemented in the CP2K package~\cite{Kuhne2020}.
In these calculations, we employed Goedecker-Teter-Hutter (GTH) pseudopotentials~\cite{Goedecker1996} to describe core-valence interactions, using molecularly optimized (MOLOPT) double-$\zeta$ polarized basis sets for Bi, O, and V, and triple-$\zeta$ basis sets for H atoms~\cite{VandeVondele2007}.
A cutoff of 600~Ry was used to expand the electron density in plane waves.

The same computational setup (CP2K with PBE0) was used to perform Born-Oppenheimer molecular dynamics (MD) simulations in the canonical ($NVT$) ensemble for five different aqueous \ce{BiVO4}(010)-water interfaces, containing 56 water molecules and approximately 150 to 200 of \ce{BiVO4} atoms.
Each simulation box had a water layer at least 15~{\AA} thick, matching the experimental water density (see Figure S4).
The initial water configuration was adopted from the previous theoretical study on the \ce{BiVO4}(010)-water interface~\cite{Wiktor2019}.
The simulations were run for 7.5~ps with a time step of 0.5~fs.
\textit{Ab initio} MD (\textit{ai}MD) simulations employed the D3 method~\cite{Grimme2010} to account for van der Waals interactions, with the temperature set at 350~K to ensure proper diffusive motion of the liquid water.
Geometric properties of the aqueous interface along the \textit{ai}MD trajectories were analyzed using the neighbor module in the Atomic Simulation Environment (ASE)~\cite{Larsen2017}.

\subsection{Training of the MLIP with Global Optimization.}

The GAP was employed to accurately represent the high-dimensional PES of the non-stoichiometric \ce{BiVO4}(010) surface system~\cite{Bartok2010,Deringer2019,Deringer2021}.
The GAP used in this study includes both two-body (2B) and many-body (MB) representations based on the Smooth Overlap of Atomic Positions (SOAP) framework~\cite{Bartok2013}.
The 2B term captures pairwise interactions between two atoms, while the MB term accounts for more complex interactions involving multiple atoms within the local atomic environments of the system.
The hyperparameters during the fitting process were carefully chosen and are listed in Table~S1.

The initial GAP model was trained on a minimal set of structures and was further refined by employing an active learning protocol, which integrates the generation of new training data with the actual surface exploration process.
Details of the active learning protocol and the quality assessment of the resulting GAP are discussed in the Supporting Information (SI).

The global geometry optimization using the GAP was conducted through a molecular dynamics (MD)-based simulated annealing (SA) protocol.
This process was carried out using LAMMPS~\cite{Plimpton1995}, employing the velocity Verlet algorithm~\cite{Swope1982} with a time step of 1~fs.
The simulations were performed within a canonical ($NVT$) ensemble, where temperature control was maintained using a Berendsen thermostat~\cite{Berendsen1984}.

In the SA protocol, the system was gradually heated from 200 to 800~K and subsequently cooled back to 200~K at a constant rate of 1.2~K/ps.
Following each SA cycle, the resulting structure underwent full geometry optimization using conjugate gradient minimization, applying the same convergence criteria as used in the DFT calculations.

A harmonic repulsive potential was applied to the slab model during heating to prevent the desorption of molecular species from the surface during the high-temperature process.
A zero potential was applied at the topmost layer of the surface slab, which was then linearly increased to a maximum value of 10~eV at a height of 10~{\AA} above the surface.
During cooling, the potential was gradually reduced and removed before the geometry optimization began, ensuring that it did not affect the final structure.

\section{RESULTS AND DISCUSSION}
\subsubsection{Global Optimization for Reconstructed \ce{BiVO4}(010) Sampling}

\begin{figure}[t!]
\center
\includegraphics[width=0.65\textwidth]{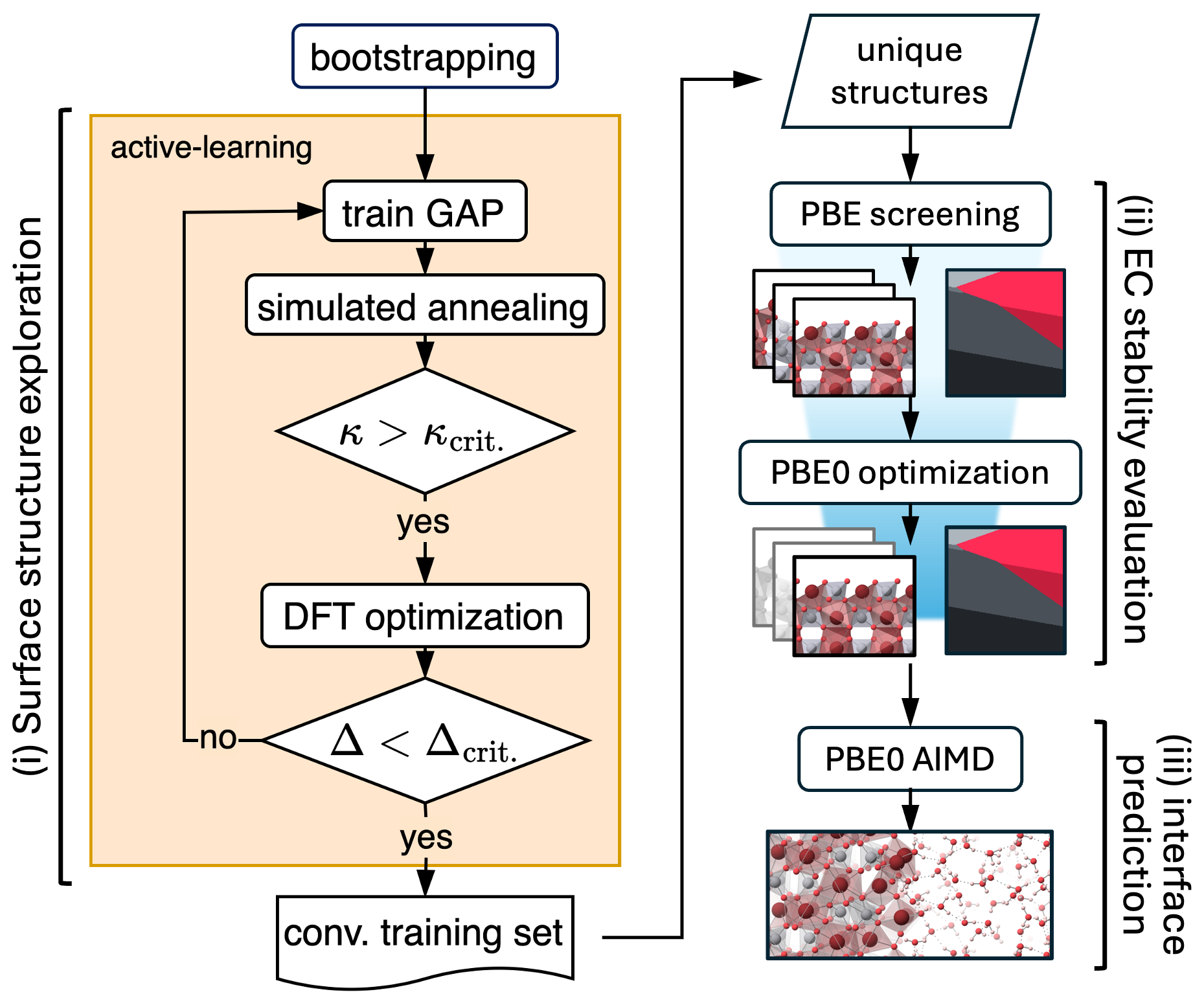}
\caption{Computational workflow for structure exploration, stability evaluations of reconstructed \ce{BiVO4}(010) surfaces, and prediction of dynamical properties at the aqueous interfaces of \ce{BiVO4}(010).}
\label{fig1}
\end{figure}

To circumvent the high computational costs associated with first-principles calculations for surface structure sampling, we employed a low-cost MLIP surrogate model, trained with a finite set of DFT data.
Specifically, we utilized GAP within an active-learning workflow that integrates the creation of new training data with the surface exploration process \cite{Timmermann2021} (see Figure~\ref{fig1} and SI for details on the active learning workflow and the simulation procedure).

This workflow begins with a minimal initial set (total 65) of Bi-V-O training structures (see SI for details).
Multiple canonical global geometry optimization runs are then performed via simulated annealing, covering a broad range of \ce{BiVO4} surface stoichiometries within the $p(1\times1)$ surface unit cell, driven by the preliminary GAP.
Notably, our simulated annealing procedure allows lateral mass transfer of surface species at high temperatures, leading to significant structural rearrangements in the surface and sub-surface regions (up to approximately 5-10~{\AA} from the outermost layer).
Meanwhile, the inner regions retain their bulk-like bonding characteristics, which is highly suited for observing surface reconstruction behavior.
The globally optimized configurations that are sufficiently distant of similarity ($\kappa > \kappa_{\rm crit}$, see SI and Eq.~S2) are added to the training set following constrained local DFT optimization.
This process is iteratively repeated until either no new unknown structures are identified based on $\kappa_{\rm crit}$, or the out-of-sample force error for the most recently added structures falls below a specified threshold.

To explore the compositional space of the \ce{BiVO4}(010) surface, we considered 13 different compositions, including variations in surface Bi/V cation ratios and oxygen content.
These surface compositions were initially generated by truncating the optimized bulk geometry and were categorized into three groups based on the relative ratios of cations in the outermost layer: Bi-rich, V-rich, and Bi/V mixed surfaces.
Specifically, the Bi-rich surface structures contain only Bi cations in the outermost layer, while the V-rich structures contain only V cations.
The Bi/V mixed surfaces maintain the same Bi/V ratio as the stoichiometric surface.
For each group, the oxygen content in the outermost metal-oxygen layer was varied, allowing us to explore both reduced surfaces and highly oxidized surfaces with excess oxygen, relevant to PEC anodic conditions.

These surfaces are denoted as \textbf{t}-\ce{Bi_xV_yO_z}, reflecting the relative ratios of atoms in the outermost layers: Bi-rich surfaces (\textbf{t}-\ce{BiO2} to \textbf{t}-\ce{BiO6}), V-rich surfaces (\textbf{t}-\ce{VO2} to \textbf{t}-\ce{VO4}), and Bi/V mixed surfaces (\textbf{t}-\ce{BiVO2} to \textbf{t}-\ce{BiVO6}).
During the sampling stage, surface adsorbates, such as hydroxyl groups, were not considered to avoid the complexities associated with water dissociation under PEC conditions.

Briefly, the active learning process resulted in only 355 DFT calculations, including a total of 7100 single-point calculations performed for the constrained geometry optimizations (20 ionic steps for each DFT calculation) of new training structures during these refinements, all completed within 3929.3 node hours ($\approx$ 164 days).
The workflow was terminated when the out-of-sample error converged below the threshold, resulting in a final training set containing 631 structures with 111765 force components.
Consequently, a total of 494 unique surface configurations within the $p(1 \times 1)$ unit cell were identified by comparing their structural similarities $\kappa$.

\subsubsection{Electrochemical Stability of Reconstructed \ce{BiVO4}(010)}

\begin{figure*}[t!]
\center
\includegraphics[width=1.0\textwidth]{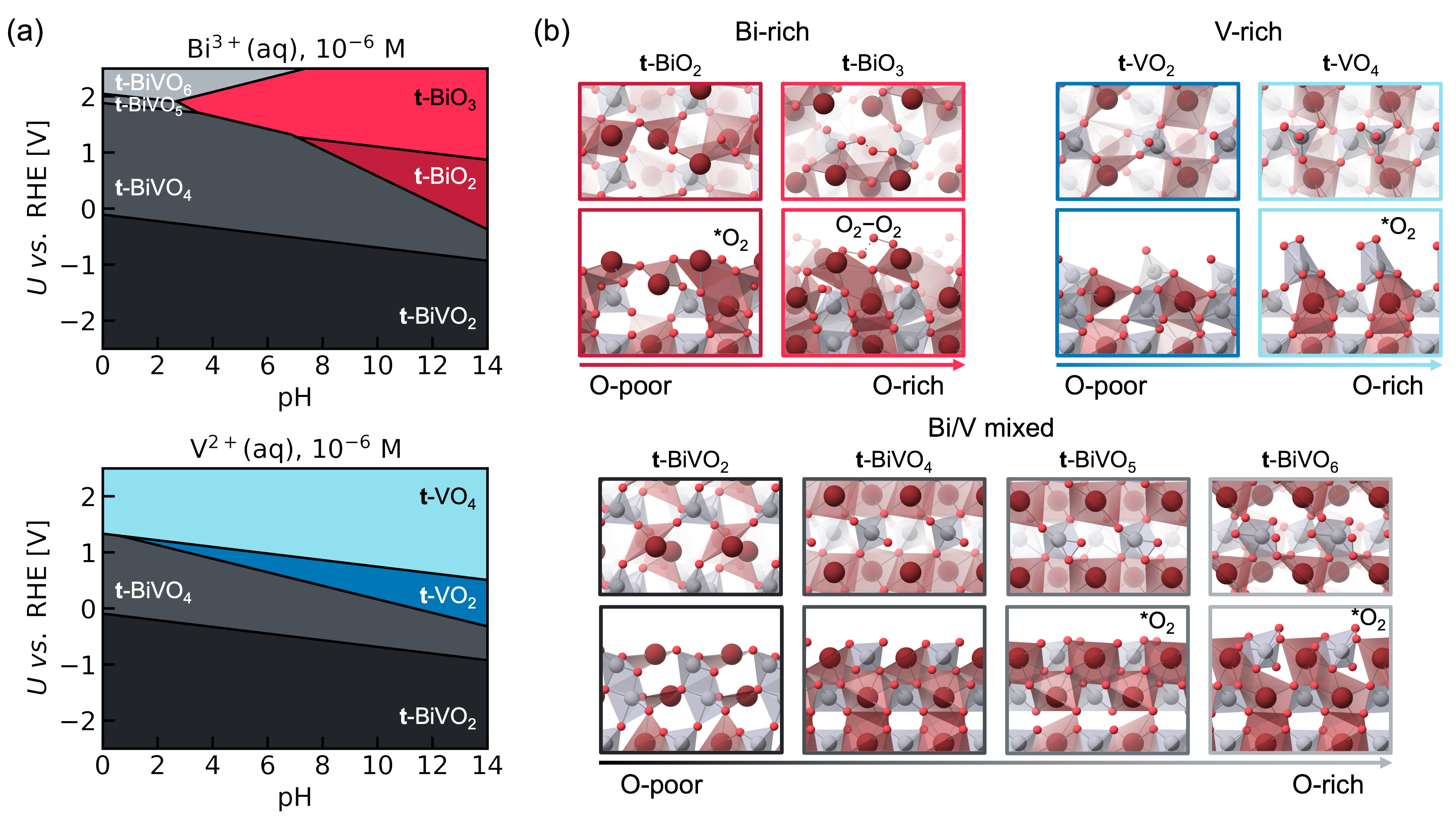}
\caption{(a) PBE0 surface Pourbaix diagrams for different \ce{BiVO4}(010) surfaces as a function of pH and potential ($U$ vs. RHE) in  aqueous \ce{Bi^{3+}} and \ce{V^{2+}}-rich electrolytes (concentration: 10$^{-6}$~M). (b) Atomic structures of Bi-rich, V-rich, and Bi/V mixed surfaces. Bi, V, and O atoms are represented by purple, gray, and red spheres, respectively. \ce{BiO_n} and \ce{VO_n} was highlighted with colored polyhedrons.}
\label{fig2}
\end{figure*}

The relative EC stabilities of the structures identified during the surface structure exploration stage were evaluated by constructing a surface Pourbaix diagram at different electrode potentials $(U_{\rm RHE})$ and pH values (see the Method section and SI for the detailed EC thermodynamics framework).
For non-stoichiometric surfaces, the thermodynamic reservoirs of excess Bi or V atoms were referenced to solvated ions (i.e., \ce{Bi^{3+}} or \ce{V^{2+}}), which were estimated from experimental formation energies and DFT total energies of bulk Bi or V, assuming equilibrium between the surfaces and the ionic states in the electrolyte.

To accurately predict the EC stability of the structures, we employed a two-step DFT calculation process.
First, all 494 unique surface structures were geometry-optimized using the GGA-PBE \textit{xc} functional.
We then screened the global minimum for each stoichiometry and the five lowest metastable local minima for stoichiometries that appeared in the PBE-based surface Pourbaix diagram (53 in total), which were further optimized using the dielectric-dependent PBE0 functional.
In PBE0 calculations, we extended our $p(1\times1)$ structures to the $p(2\times2)$ supercell to allow for potential reconstructions due to the higher degree of freedom.

Figure~\ref{fig2}a presents the PBE0 surface Pourbaix diagram under two different EC environments, assuming the surfaces are in equilibrium with electrolytes containing 10$^{-6}$~M \ce{Bi^{3+}}(aq) or \ce{V^{2+}}(aq) ionic species.
Our surface prediction at the zero potential ($U_{\rm RHE}=0$~V) reveals that, regardless of the electrolyte conditions, the stoichiometric surface (\textbf{t}-\ce{BiVO4}) remains electrochemically stable in near-neutral and acidic environments.
In contrast, under basic conditions, either the Bi-rich or V-rich surfaces possibly become dominant, depending on the specific electrolyte condition.

We then focus on conditions near $\mathrm{pH} = 7$, as this is the pH at which most experiments are conducted \cite{Rettie2013,Zhong2015,Chang2015,Iwase2016,Grigioni2018,Selim2019,Meng2023}.
When the stoichiometric \textbf{t}-\ce{BiVO4} is exposed to a \ce{Bi^{3+}} ion electrolyte, the outermost layer reconstructs into Bi-rich structures (e.g., \textbf{t}-\ce{BiO2} or \textbf{t}-\ce{BiO3}) as $U_{\rm RHE}$ increases.
Specifically, \textbf{t}-\ce{BiVO4} transitions to \textbf{t}-\ce{BiO2} at $+0.91$~V and then to \textbf{t}-\ce{BiO3} at $+1.18$~V.
In a \ce{V^{2+}} ion electrolyte, V-rich surface structures such as \textbf{t}-\ce{VO2} and \textbf{t}-\ce{VO4} begin to form as $U_{\rm RHE}$ increases, with \textbf{t}-\ce{VO2} appearing around $+0.29$~V, followed by a transition to \textbf{t}-\ce{VO4} at $+0.82$~V.

It is important to consider how our surface stability predictions, obtained under EC conditions, can be extended to PEC conditions.
A recent experiment reported that the PEC onset potential of \ce{BiVO4} was $U \simeq +0.24$~V vs. RHE, with the PEC onset negatively shifted by 1.7~V relative to the EC onset due to the photovoltage generated by illumination~\cite{Li2024}. In this regard, the experimentally observed surface in the dark is likely stoichiometric \textbf{t}-\ce{BiVO4} at zero potential.
More importantly, under this simple assumption, we conclude that \textbf{t}-\ce{BiO2} and \textbf{t}-\ce{BiO3} surfaces could be experimentally observed as Bi-rich surfaces during photocharging and anodic conditions. 

\subsubsection{Structures of Reconstructed \ce{BiVO4}(010)}
We now turn our focus to the structural properties of electrochemically stable surfaces.
The surface structures identified in the phase diagram are illustrated in Figure~\ref{fig2}b.
We begin by describing the stoichiometric \textbf{t}-\ce{BiVO4} surface, where the outermost layer consists of distorted \ce{BiO5} polyhedra and \ce{VO4} tetrahedra.
Our predictions of surface geometry show a deviation from a previous report, which identified an outermost Bi coordination number of 6 with all Bi-O bonds less than 2.55~{\AA}~\cite{Wang2020}.
However, our stable surface structure possesses a \ce{BiO5} in the outermost layer, exhibiting 5 shorter Bi-O bond lengths with an average of 2.35~{\AA}, in contrast to one Bi-O distance of 2.71~{\AA}, which is not counted as a Bi-O bond.
We found the surface with \ce{BiO6} to be metastable by 0.1~meV/{\AA}$^{2}$ compared to the surface with \ce{BiO5}, suggesting that the local offset of Bi in the polyhedron contributes to the surface stability.

The outermost layers of the reconstructed \textbf{t}-\ce{BiO2} and \textbf{t}-\ce{BiO3} surfaces undergo significant structural rearrangements compared to the stoichiometric \textbf{t}-\ce{BiVO4} (see Figure~\ref{fig2}b).
These layers are characterized by low-coordinated (highly distorted) \ce{BiO_n} ($n = 3, 4$) polyhedra, with shortened average Bi-O bond lengths of $2.22 \pm 0.15$~{\AA} for \textbf{t}-\ce{BiO2} and $2.23 \pm 0.14$~{\AA} for \textbf{t}-\ce{BiO3}, compared to the corresponding bulk bond length of 2.45~{\AA}. 
Interestingly, in both \textbf{t}-\ce{BiO2} and \textbf{t}-\ce{BiO3}, a bare (under-coordinated) Bi atom is exposed to the vacuum, as shown in Figure~\ref{fig2}b.
Furthermore, the contraction of the Bi-O bonds in the outermost layer induces additional distortions in the subsurface \ce{BiO_n} ($n = 5, 6, 7$) polyhedra, deviating from the defect-free configuration where the subsurface Bi of the stoichiometric surface typically has a coordination number of 8.

The outermost layers of both \textbf{t}-\ce{VO2} and the metastable \textbf{t}-\ce{VO3} which is only slightly less favorable than \textbf{t}-\ce{VO4} by 16~meV/{\AA} at pH$=7$ and $U_{\rm RHE}=1$~V, as a structural derivative of \textbf{t}-\ce{VO4} (see Figure~S3) are also highly corrugated, exposing \ce{VO4} tetrahedra to the vacuum.
Unlike the Bi-rich surfaces, the \ce{VO4} tetrahedral motifs in \textbf{t}-\ce{VO2} and \textbf{t}-\ce{VO3} are preserved in both the surface and subsurface regions, similar to the stoichiometric surface.
This observation aligns with the general bonding characteristics of V oxides, such as \ce{V2O5}~\cite{Mosset1982} and \ce{VO2}~\cite{Wagner2021}, where tetrahedral \ce{VO4} units serve as structural building blocks.
However, their connectivity was rather different from that of the stoichiometric surface, showing edge-sharing of outermost \ce{VO4} with subsurface \ce{VO4}.
The \textbf{t}-\ce{VO4} surface, stable at high $U_{\rm RHE}$, can be understood as additional oxygen adsorption at the tetragonal \ce{VO4} of \textbf{t}-\ce{VO3}, forming \ce{VO5} (see Figure~\ref{fig2}b).

Interestingly, the outermost structural units on \textbf{t}-\ce{BiO2} and \textbf{t}-\ce{VO4} surfaces possess \ce{O2}-like bonding, where O atoms in the outermost \ce{BiO4} and \ce{VO5} polyhedra move closer ($d_\text{O-O}$ of 1.49 and 1.32~{\AA}, respectively), resembling the bond length of the \ce{O2} molecule (1.21~{\AA}).
This \ce{O2}-like bonding within \ce{BiO_n} and \ce{VO_n} polyhedra was also observed on highly oxidized mixed surfaces at low pH conditions (around $U_{\rm RHE}$ of 2.00~V) under Bi-rich environments (see \textbf{t}-\ce{BiVO5} and \textbf{t}-\ce{BiVO6} in Figure~\ref{fig2}b).
The presence of \ce{O2}-like motifs is not surprising, as they have been observed in defect-free bulk, at the (010) surface, and even at the \ce{BiVO4}(010)-water interface~\cite{Ambrosio2019}, suggesting that the \ce{O2}-like motif could lead to a water dissociation.

Another significant feature was observed on the \textbf{t}-\ce{BiO3} surface: \ce{O2} dimerization ($d_\text{O-O}$ of 1.39~{\AA}) occurs as a dangling oxygen atom adsorbs onto the outermost \ce{BiO3} and \ce{BiO4} polyhedra.
This behavior is attributed to the tendency to maintain lower Bi coordination at the surface, even under highly oxidized conditions.
Notably, two dangling $^{*}$\ce{O2} molecules on neighboring \ce{BiO3} and \ce{BiO4} units interact, forming \ce{O2}-\ce{O2} bonds with a distance of 1.54~{\AA}.
This \ce{O2}-\ce{O2}-like structural motif has been proposed as an important transition state in the oxygen evolution reaction (OER) on the \ce{IrO2} surface \cite{Binninger2022}.

Altogether, the structural models predicted using MLIP-sampling coupled with EC thermodynamics frameworks suggest that the structures of the reconstructed overlayers on \ce{BiVO4}(010) are highly complex, extending beyond the bond features of bulk-truncated surfaces.
In particular, those surfaces exhibit the complicated connectivity between low-coordinated \ce{BiO_n} ($n = 3, 4$) and \ce{VO_n} ($n = 4, 5$) polyhedra with significant bonding distortions for Bi-rich surfaces.
It is also important to note that the Bi-rich and V-rich surface models proposed in previous theoretical work \cite{Lee2021}, which possess the same outermost stoichiometries as our \textbf{t}-\ce{BiO2} and \textbf{t}-\ce{VO2} surfaces, do not appear in our Pourbaix diagram, indicating that our globally optimized predicted surfaces are more stable than the chemical-intuition-driven structures.

\subsubsection{Aqueous \ce{BiVO4}(010) Interfaces and their Dynamical Properties}

\begin{figure*}[t!]
\center
\includegraphics[width=1.0\textwidth]{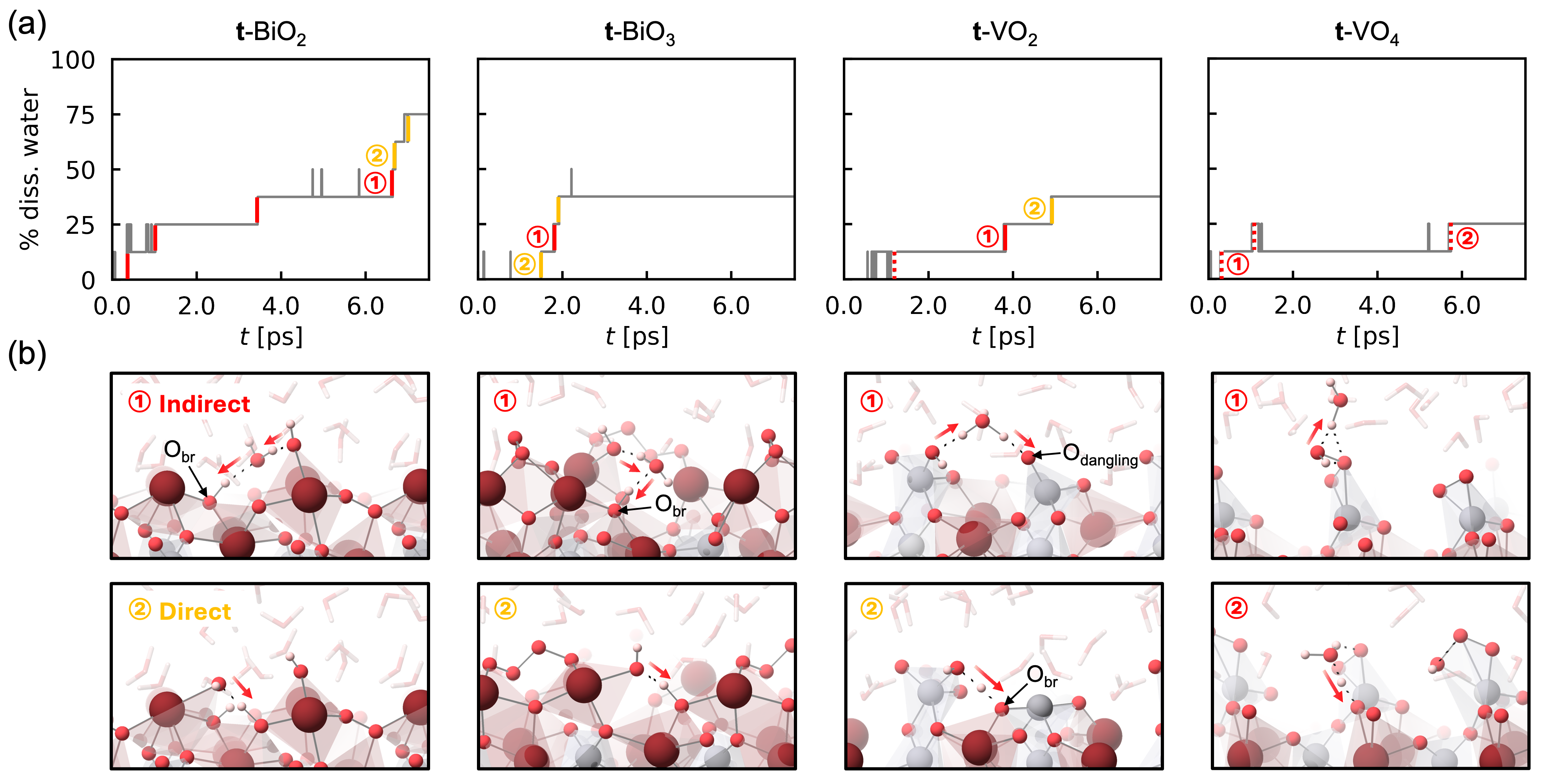}
\caption{Water dissociation on reconstructed \ce{BiVO4} surfaces (\textbf{t}-\ce{BiO2}, \textbf{t}-\ce{BiO3}, \textbf{t}-\ce{VO2}, and \textbf{t}-\ce{VO4}). (a) Percentage of dissociated water molecules over time. (b) Snapshots of interfaces, highlighting the indirect and direct water dissociation processes at the time marked in (a). The steps in (a) represent individual water dissociation events, with red and yellow steps indicating indirect and direct dissociation processes, respectively. Bi, V, O, and H atoms are represented by purple, gray, red, and white spheres, respectively.}
\label{fig3}
\end{figure*}

To understand the dynamical properties of aqueous (electrochemically stable) \ce{BiVO4} surfaces (\textbf{t}-\ce{BiVO4}, \textbf{t}-\ce{BiO2}, \textbf{t}-\ce{BiO3}, \textbf{t}-\ce{VO2}, \textbf{t}-\ce{VO4}), we conducted 7.5~ps long \textit{ai}MD simulations with at least a 15~{\AA} thick water layer (containing 56 water molecules) using the PBE0 functional. 

While many theoretical studies have broadly explored the aqueous interfacial properties of \ce{BiVO4}, none, to the best of our knowledge, have observed spontaneous water dissociation on \ce{BiVO4} surfaces.
Most previous theoretical works have focused on the aqueous interfacial properties of the stoichiometric \ce{BiVO4}(010) or (hydroxylated) Bi-rich (010) surfaces driven by chemical intuition.
Agreeing with these studies, our simulations found no dissociative water on the stoichiometric surface and its derivative with excess O atoms, i.e., \textbf{t}-\ce{BiVO6}.
However, the reconstructed Bi- and V-rich surfaces (\textbf{t}-\ce{BiO2}, \textbf{t}-\ce{BiO3}, \textbf{t}-\ce{VO2}, \textbf{t}-\ce{VO4}) exhibited significant surface hydration by spontaneous water dissociation, at the early stages of the PBE0-MD simulations.

Figure~\ref{fig3}a presents the time evolution plots of the water dissociation at each aqueous interface, along with snapshots in Figure~\ref{fig3}b illustrating the hydration process.
The dissociation percentage was defined as the number of dissociated water molecules with respect to the total number of outermost metal atoms.
The steps in Figure~\ref{fig3}a represent individual water dissociation events, with red and yellow steps indicating indirect and direct dissociation processes, respectively, while spikes correspond to spontaneous water dissociation followed by recombination as a resonance state.

The dissociation fractions ($^*$OH and $^*$H) increase over time on the reconstructed surfaces (see Figure~S5), Bi-rich surfaces (\textbf{t}-\ce{BiO2} and \textbf{t}-\ce{BiO3}) exhibit a higher degree of water dissociation with a maximum dissociation percentage of 75~{\%} compared to V-rich surfaces (\textbf{t}-\ce{VO2} and \textbf{t}-\ce{VO4}) (see Figure~\ref{fig3}a). 

The detailed dissociation mechanism on Bi-rich surfaces is further elaborated in Figure~\ref{fig3}b.
In the indirect water dissociation mechanism, an adsorbed water molecule at the bare Bi site ($^*$\ce{H2O}) donates a proton to an adjacent water molecule, forming a hydronium ion (\ce{H3O+}).
This intermediate \ce{H3O+} then transfers its excess proton to the bridging oxygen (O$_{\rm br}$) between two edge- or corner-shared \ce{BiO_n} units.
This process results in the formation of $^*$OH at the bare surface Bi (i.e., \ce{BiO3} or \ce{BiO4}) and $^*$H at O$_{\rm br}$.
The direct mechanism similarly includes proton transfer from $^*$\ce{H2O} at an under-coordinated Bi site to O$_{\rm br}$, but without an intermediate water molecule. 

On the \textbf{t}-\ce{VO2} surface, a water molecule adsorbed at a \ce{VO4} unit can transfer a proton to an adjacent water molecule, which then passes it to the outermost dangling oxygen on the neighboring \ce{VO4} unit, or directly to the O$_{\rm br}$ between the \ce{VO4} and a subsurface \ce{BO5} unit.
The red dashed step indicates indirect dissociation events where the intermediate \ce{H3O+} ion does not immediately transfer its proton to the surface oxygen but instead exchanges protons with other water molecules for a significant period (refer to dark gray markers in Figure~S5).
Eventually, the \ce{H3O+} transfers its proton to the surface (dangling) oxygen as it moves closer, indicating less favorable proton adsorption compared to Bi-rich surfaces.
The direct water dissociation process is an even rarer event, occurring only when the dangling oxygen on the neighboring \ce{VO2} unit moves closer to the subsurface Bi atom and binds as O$_{\text{br}}$.
Similarly, the indirect mechanism is dominant on the \textbf{t}-\ce{VO4} surface, where a water molecule adsorbed at the terminal oxygen of the \ce{VO5} unit transfers a proton to a water molecule above, as shown in Figure~\ref{fig3}b.
The \ce{H3O+} diffuses by delivering the proton and eventually transfers it to the O$_{\rm br}$ between the \ce{VO5} and the subsurface \ce{BiO6}.

\subsubsection{Water Dissociation Mechanism of Bi-rich Surface}

\begin{figure*}[t!]
\center
\includegraphics[width=1.0\textwidth]{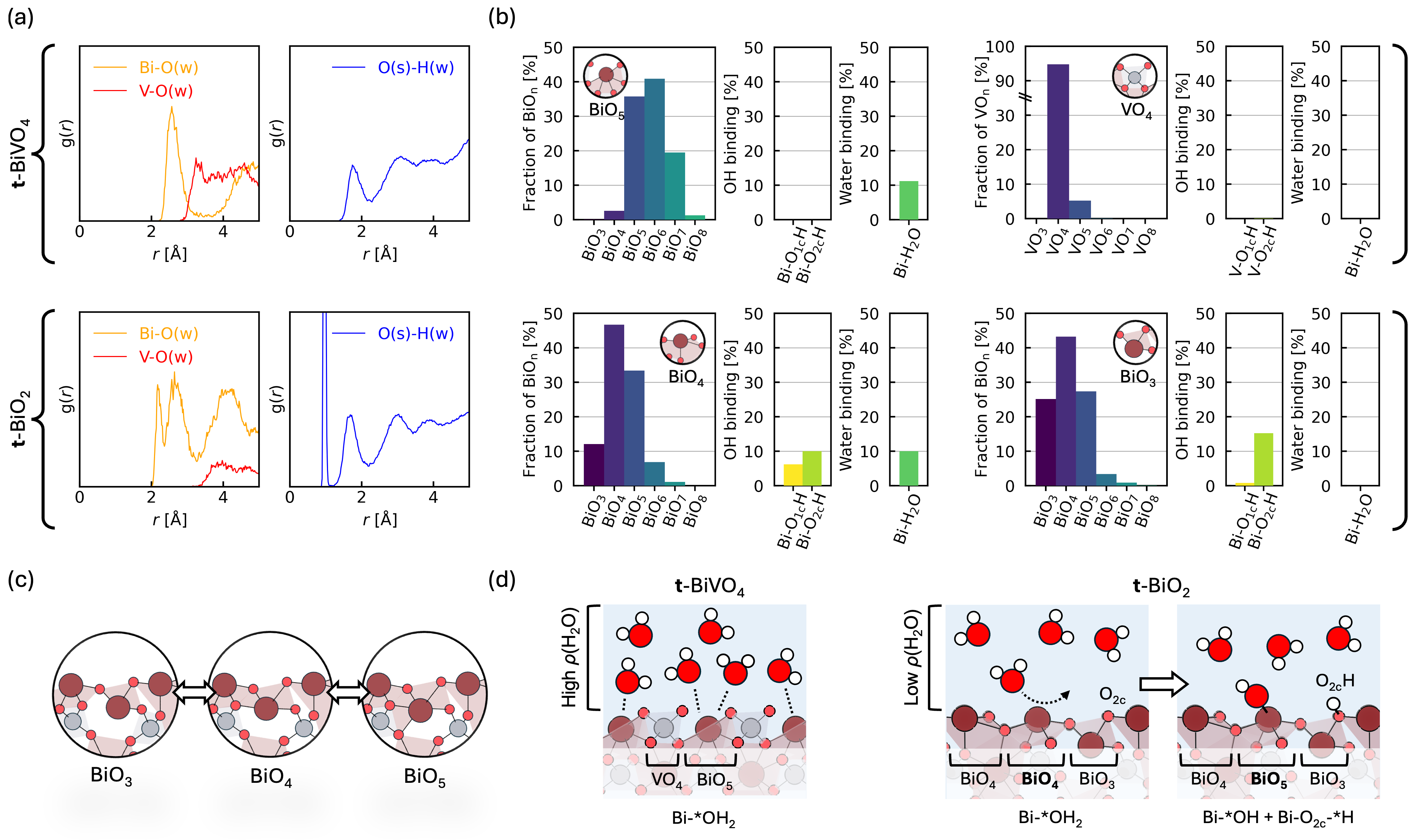}
\caption{(a) Radial distribution functions $g(r)$ for selected surface atoms on the \textbf{t}-\ce{BiVO4} and \textbf{t}-\ce{BiO2} surfaces. (b) Fraction of surface \ce{MO_n} polyhedra and percentage of $^*$OH and $^*$H$_2$O surface adsorbates amongst oxygen species bonded with the outermost Bi and V atoms. Schematic illustrations of (c) the structural perturbation of surface \ce{BiO_n} polyhedron, and (d) the water adsorption/dissociation mechanism on \textbf{t}-\ce{BiVO4} and \textbf{t}-\ce{BiO2} surfaces.}
\label{fig4}
\end{figure*}

To further investigate the dissociation mechanism of \textbf{t}-\ce{BiO2}, which is a potential surface for experimental observation under photocharging conditions and exhibits significant water dissociation, we first present the radial distribution function $g(r)$ of selected atoms in Figure~\ref{fig4}a, compared with the stoichiometric \textbf{t}-\ce{BiVO4} surface.
The minor peak at approximately 2.18~{\AA} in the RDF of O atoms in the water layer (denoted as O(w)) and surface Bi atoms indicates the presence of dissociative water, while the major peak at 2.64~{\AA} corresponds to molecular water adsorption, observed in both \textbf{t}-\ce{BiO2} and \textbf{t}-\ce{BiVO4}.
Further evidence of dissociative water can be seen in the RDF of surface oxygen (O(s)) and hydrogen in water (H(w)), where the peak at approximately 0.98~{\AA} is attributed to $^*$OH at the bare surface Bi, a feature absent on the \textbf{t}-\ce{BiVO4} surface.

Notably, despite the significant water dissociation observed on the \textbf{t}-\ce{BiO2} surface, the water density near this surface is relatively low compared to the \textbf{t}-\ce{BiVO4} surface (see Figure~S4), suggesting that the reconstructed surfaces exhibit lower hydrophilicity.
This reduced hydrophilicity may be attributed to the relatively high degree of surface corrugation and the presence of $^*$OH at the surface.

To understand the impact of molecular water adsorption and water dissociation on surface geometries, the coordination of the outermost Bi atoms is obtained from MD trajectories and plotted in Figure~\ref{fig4}b.
We observe that water adsorption and hydration processes induce geometric perturbations in the \ce{BiO_n} polyhedra on the surface.
In particular, the coordination number distribution of the outermost Bi (with a Bi-O bond cutoff of 2.65~\AA) on \textbf{t}-\ce{BiVO4} shows a variation from \ce{BiO5} to \ce{BiO7}, with \ce{BiO5} and \ce{BiO6} accounting for 75~{\%} of the population, while about 95~{\%} of \ce{VO4} units maintain their geometry.
Similarly, outermost \ce{BiO4} and \ce{BiO3} polyhedra on the \textbf{t}-\ce{BiO2} surface undergo geometric perturbations, changing the coordination environments of Bi atom (see \ref{fig4}c for schematic illustration).
Specifically, the surface \ce{BiO4} fluctuates mostly between \ce{BiO4} and \ce{BiO5}, making up approximately 80~\% of the population, while the \ce{BiO3} varies between \ce{BiO3} and \ce{BiO5} with over 95~\% of the population. 

Next, we analyze the coordination environments of oxygen species bonded with the outermost Bi or V atoms to identify the preferred water binding and spontaneous water dissociation sites.
Their distribution is also displayed in Figure~\ref{fig4}b, noting that the majority of oxygen atoms are lattice oxygen at the surface (which is excluded in our plots), with a minority assigned to molecular adsorbates such as $^*$OH and $^*$\ce{H2O}.

Water adsorption on \textbf{t}-\ce{BiVO4} predominantly occurs at the Bi site, leaving the V site unoccupied, with no $^*$\ce{H2O} binding at the V site. For \textbf{t}-\ce{BiO2}, water molecules bind to the bare Bi sites of \ce{BiO4}, while the Bi site of \ce{BiO3}, where the bare Bi atom is not exposed to the water layer, remains inert to water adsorption. The adsorbed water molecules on the bare Bi sites in \textbf{t}-\ce{BiO2} further dissociate, forming Bi-\ce{O_{1c}}H (one-fold coordinated OH) and Bi-\ce{O_{2c}H} (two-fold coordinated OH) simultaneously. In contrast, the oxygen species bonded to the Bi atom in \ce{BiO3} exhibit only a two-fold coordinated Bi-\ce{O_{2c}H} occupation, due to proton transfer resulting from water dissociation occurring at \ce{BiO4} sites.

To conclude, the mechanisms of water adsorption/dissociation on stoichiometric and Bi-rich surfaces are illustrated in the schematic in Figure~\ref{fig4}d.
Our findings emphasize the crucial role of the Bi site, rather than the V site, in water adsorption on both stoichiometric and reconstructed surfaces.
More importantly, spontaneous water dissociation occurs to a significant extent on reconstructed Bi-rich \ce{BiVO4}(010) surfaces, particularly when the surface Bi site is under-coordinated and the bare Bi is exposed to water layers.
In detail, the dissociation of water molecules at the Bi site leads to the formation of $^*$OH, while the released protons in the water layer tend to occupy the O$_{\rm br}$ site through \ce{H3O+} transfers.

\section{CONCLUSION}
We employed a GAP within an active-learning framework to investigate \ce{BiVO4}(010) surface reconstructions recently observed during the PEC anodic cycle. By integrating an MLIP active-learning workflow with simulated annealing, designed to predict significant structural rearrangements across configurational spaces, we identified 494 unique surface structures. We then evaluated the relative electrochemical stability of these surfaces under Bi- and V-rich conditions using DFT and dielectric-dependent hybrid functional calculations. This analysis ultimately narrowed our selection to six electrochemically stable structures that could potentially be observed experimentally.

We subsequently conducted \textit{ab initio} molecular dynamics simulations to investigate the aqueous interfaces of selected reconstructed \ce{BiVO4}(010) surfaces. Our simulations revealed that, unlike the stoichiometric surface, the reconstructed Bi- and V-rich surfaces exhibit significant (spontaneous) water dissociation, with Bi-rich surfaces showing the highest degree of dissociation. These findings suggest that surface reconstruction of \ce{BiVO4}(010) is essential for enhanced surface hydration, as the bare and low-coordinated Bi sites resulting from this reconstruction play a critical role as active centers for spontaneous water dissociation.

Here, we establish reliable surface structure models of \ce{BiVO4}(010) under PEC conditions through our surface structure exploration workflow, which overcomes the limitations of previous chemical-intuition-driven models. We anticipate that our predicted surface structures will be validated by future experimental studies.

This work not only lays the groundwork for further investigations into the surface structures and reactivity of \ce{BiVO4} but also offers a robust theoretical framework for enhancing our understanding of complex surface and interface processes in multicomponent compounds. Moving forward, our future work will focus on characterizing the reconstructed \ce{BiVO4}(010)-water interfaces from an electronic structure perspective, including the interface band offset, which is critical for understanding PEC anodic reactivity.

\section{Data Availability}
All \textit{ab initio} training data will be provided on the Materials Cloud Archive upon publication.

\section{Author Contributions}
Y. Lee conducted the simulations, performed the characterizations, and wrote the manuscript. T. Lee contributed to the conception and design of the simulations and wrote the manuscript.

\begin{acknowledgement}
This work is also supported by research funds for newly appointed professors of Jeonbuk National University in 2023. 
\end{acknowledgement}

\begin{suppinfo}
\textbf{Supporting Information.} 
(PDF) Theoretical details, including the force locality test, chosen GAP hyperparameters, the accuracy of the GAP model, MD and DFT simulation protocols, and methods for assessing electrochemical stability. Additionally, water density plots for aqueous interfaces of \ce{BiVO4}-water, and the time-dependent evolution of the relative z-coordinate of water dissociation products are provided.
\end{suppinfo}

\bibliography{manuscript}
\end{document}